\documentclass{iaus}
\usepackage{graphicx}
\usepackage{times}

\def\apj{\textit{ApJ}}

\def\aap{\textit{A\&A}}            
\def\mnras{\textit{MNRAS}}         
\def\pasp{\textit{MNRAS}}  

\title[ The formation of S0 Galaxies] 
{ Stellar populations in the bulges of S0s \\ and the formation of S0 galaxies}

\author[Alfonso Arag\'on-Salamanca]   
{Alfonso Arag\'on-Salamanca}

\affiliation{
School of Physics and Astronomy,
  University of Nottingham, Nottingham, NG7 2RD, UK.
 \break email: alfonso.aragon@nottingham.ac.uk}

\pubyear{2007}
\volume{245}  
\pagerange{1--4}
\date{?? and in revised form ??}
\jname{Formation and Evolution of Galaxy Bulges}
\editors{M.\ Bureau, E.\ Athanassoula, \& B.\ Barbuy, eds.}
\begin{document}

\maketitle

\begin{abstract}

The stellar populations in the bulges of S0s, together with the galaxies'
dynamics, masses and globular clusters, 
contain very interesting clues about their formation. 
I present here recent evidence suggesting that S0s are the descendants of
fading spirals whose star formation ceased.  

\end{abstract}


Lenticular, or S0, galaxies make up some 25\% of large galaxies in the
local Universe (Dressler 1980), so understanding how
they form must constitute a significant element of any explanation of
galaxy evolution.  Their location at the crossroads between
ellipticals and spirals in Hubble's tuning-fork diagram underlines
their importance in attempts to develop a unified understanding of
galaxy evolution, but also means that it is not even clear to which of
these classes of galaxy they are more closely related.

One often-cited piece of evidence comes from the fact that the
proportion of S0s is substantially smaller in distant ($z\sim0.5$) clusters
than in nearby ones, while spirals show the opposite trend (Dressler
et al.\ 1997), strongly suggesting a transformation from
one to the other.  However, even if this scenario is accepted, it does
not answer the question as to whether S0s are more closely related to
spirals or ellipticals, which is intimately connected to the mechanism
of transformation.  If the transformation simply involves a spiral
galaxy losing its gas content through ram pressure stripping (Gunn \&
Gott 1972) or ``strangulation'' (Larson et al.\ 1980), 
so ceasing star formation and fading into an S0, then
clearly S0s and spirals are closely related.  However, it is also
possible that mergers can cause such a transformation: while
equal-mass mergers between spirals create elliptical galaxies, more
minor mergers can heat the original disk of a spiral and trigger a
brief burst of star formation, using up the residual gas and leaving
an S0.  In such a merger scenario, the mechanism for creating an S0 is
much more closely related to that for the formation of ellipticals.

Clues to which mechanism is responsible are to be found in the
``archaeological record'' that can be extracted from spectral
observations of nearby S0s.  In particular, the present-day stellar
dynamics should reflect the system's origins, with the gentle gas
stripping of a spiral resulting in stellar dynamics very similar to
the progenitor spiral, while the merger process will heat the stars,
resulting in kinematics more dominated by random motions, akin to an
elliptical.  In addition, the absorption line strengths can be
interpreted through stellar population synthesis to learn about the
metallicity and star formation histories of these systems.  Even more
interestingly, these dynamical and stellar properties can be compared
to see if a consistent picture can be constructed for the formation of
each system. I present here some recent 
evidence suggesting that such a consistent picture is indeed emerging. 


\section{Evidence from the Tully-Fisher relation}

Combining published data with high-quality VLT/FORS spectroscopy of sample
of Fornax S0s (Bedregal et al.\ 2006a)  we have carried out a combined
study of the Tully-Fisher relation and the stellar populations of these
galaxies.  Despite the relatively small sample and the considerable
technical challenges involved in determining the true rotation velocity $V_{\rm
rot}$ from absorption line spectra of galaxies with significant non-rotational
support (see Mathieu et al.\ 2002), some very interesting results arise.
S0s lie systematically below the spiral galaxy Tully-Fisher relation in both
the optical and near-infrared (Figure~1). If S0s are the descendants of spiral
galaxies, this offset can be naturally interpreted as arising from the
luminosity evolution of spiral galaxies that have faded since ceasing star
formation. Moreover, the amount of fading implied by the offset of individual
S0s from the spiral relation seems to correlate with the luminosity-weighted
age of their stellar population, particularly at their centres (Figure~2). 
This correlation suggests a scenario in which the star formation clock stopped
when gas was stripped out from a spiral galaxy and it began to fade into an S0.
The stronger correlation at small radii indicates a final last-gasp burst of
star formation in this region. See Bedregal, Arag\'on-Salamanca  \& Merrifield
(2006b) for details. 

\begin{figure}
\begin{center}
\includegraphics[height=3.3in,width=4.0in,angle=0]{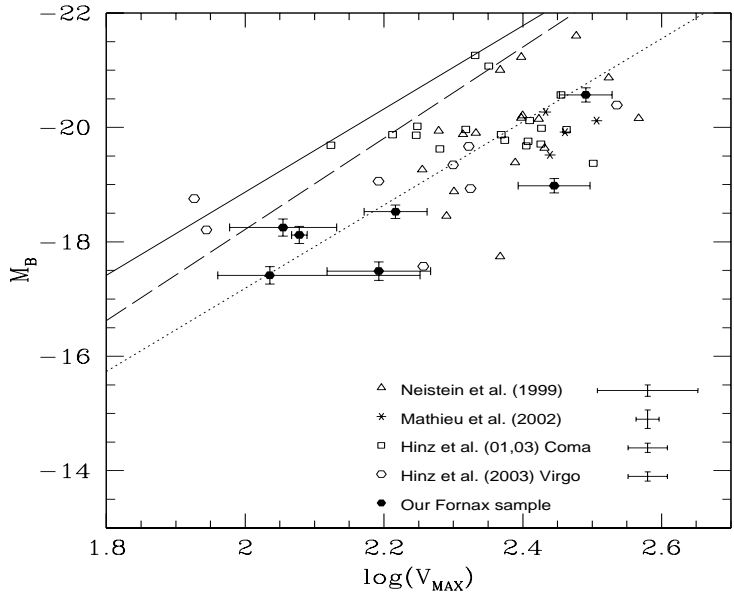}
\end{center}
\caption{$B$-band Tully-Fisher relation (TFR) for 
  S0 galaxies using
  different samples from the literature (open symbols) and our VLT Fornax   
  data (filled circles). 
  The solid and dashed lines show two independent determinations of 
  the TFR relation for local spirals. On average (dotted line),
  S0s are $\sim3$ times fainter
  than spirals at similar rotation velocities 
  (Bedregal, Arag\'on-Salamanca \& Merrifield 2006b). 
       }
\label{fig:fig1}
\end{figure}

\begin{figure}
\begin{center}
\includegraphics[height=1.65in,angle=0]{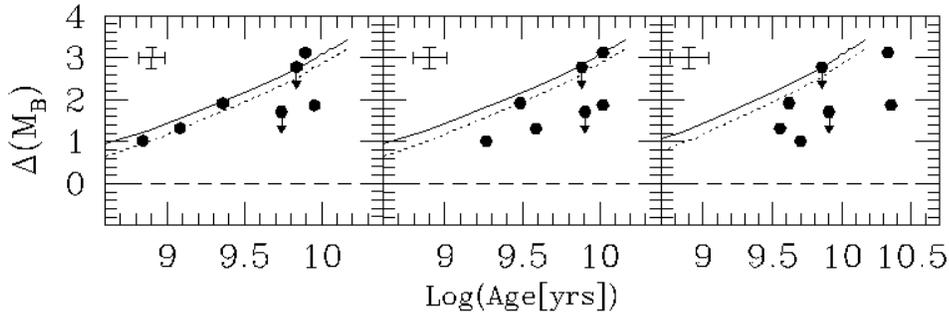}
\end{center}
\caption{
  For our VLT Fornax data we plot the
  shift in magnitudes from the
  $B$-band spiral TFR versus the stellar population age at the galaxy
  centre (left panel), at $1\,R_e$ (middle panel) and at $2\,R_e$ (right
  panel). The lines show models for fading spirals. 
  Note that the correlation is strongest for the central stellar 
  populations of the galaxies, suggesting that the last episode of star
  formation took place there (Bedregal, Arag\'on-Salamanca \& Merrifield 2006b).  
    }
\label{fig:fig2}
\end{figure}


\section{Evidence from the globular cluster populations}

Entirely consistent and independent evidence comes from our recent
studies of the properties of the globular cluster (GC) systems and stellar
populations of SOs (Arag\'n-Salamanca, Bedregal \& Merrifield 2006; Barr et
al.\ 2007).  If interactions with the intra-cluster medium are responsible  for
the transformation of spirals into S0s, the number of globular clusters in
these galaxies  will not be affected. That is probably not true if more violent
mechanisms such as galaxy-galaxy interactions are the culprit (see, e.g.,
Ashman \& Zepf 1998). If we assume that the number of globular clusters remains
constant, the GC specific frequency  ($S_N\propto\,$number of GCs per unit
$V$-band Luminosity) would increase due to the fading of the galaxy. On
average, the GC specific frequency is a factor $\sim 3$ larger for S0s than it
is for spirals (Arag\'on-Salamanca et al. 2006),  meaning that
in the process S0s become, on average,  $\sim 3$ times fainter than their
parent spiral.  Furthermore, in this scenario the amount of fading (or increase
in GC specific frequency) should grow with the time elapsed  since the star
formation ceased, i.e., with the luminosity-weighted age of the S0 stellar
population.  Figure~3 shows that this is indeed the case, adding considerable
weight to the conclusions reached from our Tully-Fisher studies.


\section{Additional evidence from the stellar populations and dynamics} 

In Bedregal et al.\ (2007) we show that the central  absorption-line indices in
S0 galaxies correlate well with the central velocity dispersions in accordance
with what previous studies found for elliptical galaxies.  However, when these
line indices are converted into stellar population properties, we find that the
observed correlations seem to be driven by systematic age and alpha-element
abundance variations, and not changes in overall metallicity as is usually
assumed for ellipticals. These correlations become even tighter when the
maximum circular velocity is used instead of the central velocity dispersion.  
This improvement in correlations is interesting because the maximum rotation
velocity is a better proxy for the S0's dynamical mass than its central
velocity dispersion. Finally,  the  $\alpha$-element over-abundance seems to be
correlated with dynamical mass, while the absorption-line-derived ages also
correlate with these over-abundances. These correlations imply that the most
massive S0s have the shortest star-formation timescales and the oldest stellar
populations,  suggesting that mass plays a large role in dictating the life
histories of S0s.


\section{Conclusions}

The stellar populations, dynamics and globular clusters of S0s  provide
evidence consistent with these galaxies being the descendants of fading spirals
whose star formation ceased.  However, caution is needed since significant
problems could still exist with this picture (see, e.g., Christlein \&
Zabludoff 2004; Boselli \& Gavazzi 2006). Moreover, the  number of 
galaxies studied
here is still small, and  it would be highly desirable to extend this kind of
studies to much larger samples covering a broad range of galaxy masses and
environments.

\begin{figure}
\begin{center}
\includegraphics[height=2.9in,angle=0]{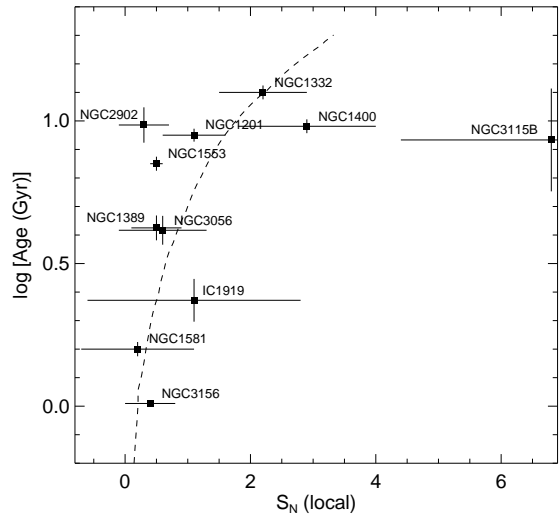}
\end{center}
\caption{
Log$_{10}$ of the luminosity-weighted ages is Gyr  
  vs.\ the globular cluster specific frequency
  ($S_N$) of S0s. The line shows the evolution 
  expected for a
  fading galaxy according to the stellar population models of 
  Bruzual \& Charlot
  (2003). The correlation between the fading of the galaxies 
  (or increase in $S_N$) and the spectroscopically-determined
  age of their stellar populations is clearly consistent with the predictions of
  a simple fading model.   
  Note that the $S_N$ value for NGC3115B
  is very unreliable and almost certainly 
  severely overestimated due
  to contamination from the GC systems of neighbouring galaxies. 
  See Barr et al.\ (2007) for details.    
    }
\label{fig:fig3}
\end{figure}

\begin{acknowledgments}

I thank A.G.\ Bedregal,  M.\ Merrifield, J.M.\ Barr, B.\ Milvang-Jensen,  S.P.\
Bamford and N. Cardiel for allowing me to discuss here results obtained with
their help.

\end{acknowledgments}


\begin{thebibliography}{}


\bibitem[Arag\'on-Salamanca, Bedregal \& Merrifield (2006)]{Aragon_et_al06}
Arag{\'o}n-Salamanca, A., Bedregal, A.~G.,  Merrifield, M.~R., 2006, 
\aap, 458. 101

\bibitem[Ashman \& Zepf(1998)]{1998gcs..book.....A} Ashman, K.~M., \& Zepf, 
S.~E.\ 1998, Globular cluster systems (Cambridge, U.K.; 
New York: Cambridge University Press, Cambridge 
astrophysics series; 30)  

\bibitem[Barr et al.(2007)]{Barr_et_al07} Barr, J.~M., Bedregal, 
A.~G., Arag{\'o}n-Salamanca, A., Merrifield, M.~R., \& Bamford, S.~P.\ 
2007, \aap, 470, 173 


\bibitem[Bedregal et al.(2006a)]{Bedregal_et_al06a} Bedregal, A.~G., 
Arag{\'o}n-Salamanca, A., Merrifield, M.~R., \& Milvang-Jensen, B.\ 2006a, 
\mnras, 371, 1912 


\bibitem[Bedregal et al.(2006)]{2006MNRAS.373.1125B} Bedregal, A.~G., 
Arag{\'o}n-Salamanca, A., \& Merrifield, M.~R.\ 2006b, \mnras, 373, 1125 

\bibitem[Bedregal et al.(2007)]{Bedregal_07} Bedregal, A.~G., 
Arag{\'o}n-Salamanca, A., Merrifield, M.~R. \& Cardiel, N.\ 2007, \mnras, 
submitted 

\bibitem[Boselli \& Gavazzi(2006)]{2006PASP..118..517B} Boselli, A., \& 
Gavazzi, G.\ 2006, \pasp, 118, 517 

\bibitem[Bruzual \& Charlot(2003)]{2003MNRAS.344.1000B} Bruzual, G., \& 
Charlot, S.\ 2003, \mnras, 344, 1000 

\bibitem[Christlein \& Zabludoff(2004)]{2004ApJ...616..192C} Christlein, 
D., \& Zabludoff, A.~I.\ 2004, \apj, 616, 192 

\bibitem[Dressler(1980)]{1980ApJ...236..351D} Dressler, A.\ 1980, \apj, 
236, 351 

\bibitem[Dressler et al.(1997)]{1997ApJ...490..577D} Dressler, A., et al.\ 
1997, \apj, 490, 577 

\bibitem[Gunn \& Gott(1972)]{1972ApJ...176....1G} Gunn, J.~E., \& Gott, 
J.~R.~I.\ 1972, \apj, 176, 1 

\bibitem[Larson et al.(1980)]{1980ApJ...237..692L} Larson, R.~B., Tinsley, 
B.~M., \& Caldwell, C.~N.\ 1980, \apj, 237, 692 

\bibitem[Mathieu et al.(2002)]{Mathieu_et_al02} Mathieu, A., 
Merrifield, M.~R., \& Kuijken, K.\ 2002, \mnras, 330, 251 
 

\end{thebibliography}
\end{document}